\documentstyle[aps,prl,twocolumn]{revtex}
\newcommand{\be}{\begin{equation}}
\newcommand{\ee}{\end{equation}}
\newcommand{\bea}{\vspace{0.25cm}\begin{eqnarray}}
\newcommand{\eea}{\end{eqnarray}}

\def\PLA{{Phys. Lett.}  A }
\def\PRL{{Phys. Rev. Lett.} }
\def\PRA{{Phys. Rev.} A }

\begin{document}
\title{Quantum Clock Synchronisation based on entangled photon pairs
transmission}
\author{M.Genovese and C. Novero }
\address{ Istituto Elettrotecnico Nazionale Galileo Ferraris, 
Str. delle Cacce 91,
I-10135 Torino, Italy
}
\maketitle

\begin{abstract}
We discuss the possibility of synchronising two atomic clocks exchanging
entangled photon pairs through a quantum channel. A proposal for
implementing practically such a scheme is discussed. 

{\bf PACS: } 03.67.-a, 03.67.Hk, 06.30.Ft, 95.55.Sh, 03.65.-w 
\end{abstract}

\vspace{0.5cm} 

In the last years the study of quantum information has led to new
perspectives in many different fields, originating relevant progresses.

Recently the concept of qubits and of their manipulation has been used for
describing an atomic clock with the purpose of obtaining new improvements
of this apparatus. In particular, the use of entangled atoms has been shown
to offer the possibility of improving the errors affecting time
determination \cite{Chiara} and a new scheme for synchronising clocks based
on entanglement has been presented \cite{Jozsa}.
For what concerns this last paper, an interesting way for a perfect
synchronisation of atomic clocks (in the sense of having a common origin of
time counting) has been proposed, using shared pairs of entangled atoms.
The main problem of this scheme is the big difficulty of sharing a set of
entangled atoms between two laboratories (in the following we will use for
them the conventional names Alice and Bob): as far as we know no practical
way of obtaining such a result has been proposed. Purpose of this paper is
to show how this entanglement can be created using entangled pairs of
photons, which are much easier to be transmitted (in effect transmissions
over more than 30 km have already been achieved \cite{lontano}).
  
A general definition of a quantum clock is a tensor product of identical
vectors in a n-dimensional Hilbert space, which are eigenstates of the
Hamiltonian operator.
At some instant a transformation acts on each vector, mapping it in a
non-stationary state. A later measurement  allows to extract the standard
of time by the evolution of the vectors themselves.
Ref. \cite{Chiara} has shown that the measurement precision can be
remarkably increased using as input the product of n eigenvectors of the
Hamiltonian which are transformed into non-stationary  non-factorisable
entangled states. However, unluckily, in the case investigated by ref.
\cite{Chiara}, i.e. for the case of a $\oplus_{k=1}^n sl_k(2)$ (qubit)
transformation, describing a Ramsey pulse, on a tensor product of n
Hamiltonian eigenvectors of a d=2 Hilbert space which describes n atoms in
the lowest energy level, decoherence strongly reduces the power of this
scheme, or at least complicates its experimental implementation (for a
description of decoherence in this case see \cite{BOTV} and ref.s therein).
It remains opened the door for finding new physical systems where evolution
can be accomplished in a decoherence free subspace \cite{paolo} or where
open loop decoherence control techniques can be implemented \cite{viola},
eliminating this problem.

Let us now review in more detail  the Ramsey method \cite
{Ramsey1969}, used for realising nowadays atomic clocks, in terms of
quantum information language. A clock is constructed
by an ensemble of two-level systems (qubits),  whose temporal evolution
determines the time standard.
According to the International System of Units (SI), the today definition of a
second is 9192631,770 periods of oscillation of the hyperfine transition
frequency for the ground-state of the $^{133}$Cs. Anyway, in principle, any
other quantum system could be used.

Let us denote the two hyperfine levels with energy eigenvalues $E_0 < E_1$
as $|0\rangle$ and $|1\rangle$ respectively. Then, let us introduce another
basis for the 2 dimensional Hilbert space
$%
|+\rangle = \frac{1}{\sqrt{2}}(|0\rangle + |1\rangle )$ and $|-\rangle = 
\frac{1}{\sqrt{2}}(|0\rangle - |1\rangle )$. 

The realisation of an atomic clock is based on the fact that
the states $|+\rangle$ and $|-\rangle$ are not stationary states, but
evolve as: 
\begin{equation}  \label{ev+}
\begin{array}{lcl}
|+(t)\rangle & = & \frac{1}{\sqrt{2}} \left( e^{-i \Omega t/2} |0\rangle +
e^{i \Omega t/2} |1\rangle \right) \\ 
|-(t)\rangle & = & \frac{1}{\sqrt{2}} \left( e^{-i \Omega t/2} |0\rangle -
e^{i \Omega t/2} |1\rangle \right)
\end{array}
\end{equation}
where $\Omega = \omega - \omega_0$ is the difference between the applied
driving frequency $\omega$ and the resonance one $\omega_0=  ( E_{1} -
E_{0} ) / \hbar $.

At a time $t=0$ we dispose of a sample of atoms in the state $|0\rangle$,
then we apply a Ramsey pulse which corresponds, in quantum information
language, to the Hadamard transform ($(\openone /2 + e_+ - e_-)\sqrt{2}$ in
terms of the generators of the algebra of sl(2)), which is defined by the
operations $|0\rangle \rightarrow |+\rangle$ and $|1\rangle \rightarrow
|-\rangle$. 
This originates an ensemble of $ |+\rangle$ states, which begin to evolve
in time.
After some delay $t$, a second Hadamard transformation is applied.
A straightforward calculation shows that
the probabilities for observing a state $|0\rangle$
or $|1\rangle$ are respectively given by 
\begin{equation}  \label{P}
P_0(t) = \frac{1}{2} \left(1 + \cos \left( \Omega t \right) \right), \hspace{%
2mm} P_1(t) = \frac{1}{2} \left(1 - \cos \left( \Omega t \right) \right)
\end{equation}
Incidentally, this simple expression corresponds to the exact result of a
more involved calculation \cite{Audoin} when $\Omega << b$, with $b= \mu_B
B / (2 \hbar)$ and where $B$ is the amplitude of the microwave magnetic
flux density applied to the interaction region. This condition is well
satisfied in experimental realisations of atomic clocks.
 
By monitoring the oscillations of $P_1(t)$ we can identify its maximum
corresponding to  $\omega = \omega_0 $. Then $\omega$ furnishes  the 
standard of time.

Let us now give a summary of the proposal of \cite{Jozsa}.
Alice and Bob share an ensemble of entangled states 
\be
| \Psi \rangle = \frac{1}{%
\sqrt{2}} \left( |0\rangle_{A} |1\rangle_{B} - |1\rangle_{A}
|0\rangle_{B}\right)
\label{psi}
\ee
where the subscripts refer to particles held by Alice
and Bob. Furthermore, Alice and Bob are able to identify the single atoms
(for example because each of them is confined in a known position in a ions
lattice), which are labelled by $n=1,2,3, \ldots $. The state (\ref{psi})
does not
evolve in time for it is invariant for the direct product of the two
unitary evolution operators of the subsystems A and B.

In order to start the clocks Alice simultaneously measures (at $t=t_0$) all of
her atoms in the basis $\{ |+\rangle, |-\rangle \}$ inducing the collapses
randomly (with equal probability $1/2$) into
one of the following states: 
\begin{equation}  \begin{array}{lcl}
|\psi^{I}\rangle & = & |+\rangle_{A}|-\rangle_{B} \\ 
|\psi^{II}\rangle & = & |-\rangle_{A}|+\rangle_{B}
\label{collasso}

\end{array}
\end{equation}
with equal probability $\frac{1}{2}$. Alice's and Bob's clocks begin to evolve
in time, in accordance with Eq. (\ref{ev+}) - both starting synchronously
at $t=t_ 0$ in Alice and Bob's shared inertial frame. Indeed Alice's
measurement is effectively equivalent to the first one-clock
Hadamard transform in the Ramsey scheme. However the result here is a
mixture of the two equally weighted sub-ensembles I and II of Eq.
\ref{collasso}. As a result of her
measurement, Alice knows the labels belonging to the subensembles I and II
but Bob is unable to distinguish them.

Then, Alice post-selects from her entire ensemble the
sub-ensemble corresponding to Type-I qubits (for example). Since the qubits
are labelled, she can communicate to Bob (using a classical channel) which
subset of his qubits are also Type-I. Bob can thus select 
his Type-I and Type-II subensembles. Selecting the Type-II subensemble,
he will have a clock ensemble exactly in phase with a Type-I clock that
Alice started at some initial time $t_0$.
Of course some noise due to decoherence effects cannot be excluded, leading
to  a finite precision of the synchronisation scheme. An accurate study of
this effect is beyond the purpose of this paper.
 
As hinted before, the main difficulty of this scheme is that sharing an
ensemble of entangled atoms is a extremely difficult task. Practically,
there is little hope of transporting atoms to a far laboratory keeping them
entangled with atoms in the original lab. 

In order to overcome this severe difficulty one can think of using
entangled photon pairs in order to establish entanglement between two
distant atoms initially uncorrelated. Furthermore, this procedure allows to
create entanglement every time it is necessary just using a quantum line.
A similar idea has recently been proposed by ref. \cite{seth} in order to
create a quantum memory based on rubidium atoms entangled by use of a
quantum transmission of entangled photons. In the following, we will adapt
this scheme for  synchronisation.
  
In the following we focus on the "clock" transition of cesium atoms. 
The scheme consists in transmitting to Alice and Bob one member of an
entangled pair of photons both with a wave length of 852.1 nm (which
corresponds to the $D_2$ transition of cesium)
\begin{equation} 
\vert \psi \rangle = {\frac{ \vert R\rangle \vert L \rangle +  \vert L
\rangle \vert R \rangle }{\sqrt {2}}}  \label{Psi}
\end{equation}
where R and L denotes right and left circular polarisation respectively.
These entangled pairs can be easily generated using parametric down
conversion \cite{nos}.
 
Then Alice and Bob address the received photon on a specific atom which is
prepared in the ground state $6 ^2S_{1/2}$ $F=3$ ($F$ is the total angular
momentum of the nucleus plus the electronic system) $m_F=0$ state. The
photon excites the transition to 
the states $6 ^2P_{3/2}$ $F=3$ $m_F= 1$ or $m_F=-1$ according if the
received photon has a right or left polarisation. Finally, two Raman $\pi$
pulses 
transfer the state $\vert 6 ^2P_{3/2}, F=3, m_F=1 \rangle$ into $\vert 6
^2S_{1/2}, F=3, m_F=0 \rangle$ and the state $\vert 6 ^2P_{3/2}, F=3,
m_F=-1 \rangle$ into $\vert 6 ^2S_{1/2}, F=4, m_F=0 \rangle $ respectively,
where the two Zeeman levels are opportunely split thanks to an external
magnetic field ($B$). This last step puts the atom in a superposition of
the $6 ^2S_{1/2}$ $F=3$ $m_F=0$ ($ |0\rangle$) and $F=4$ $m_F=0$ ($
|1\rangle$); 
thus the desired entangled state $\frac{1}{%
\sqrt{2}} \left( |0\rangle_{A} |1\rangle_{B} - |1\rangle_{A}
|0\rangle_{B}\right)$
of atoms is created.

The transition between the states $6 ^2S_{1/2}$ $F=3$ $m_F=0$  and $F=4$
$m_F=0$ 
is known as "clock transition", for it depends only at the second order by
the external magnetic field ($\nu [F=4, m_F=0 \rightarrow F=3, m_F=0]
=\nu_0 + 427 \cdot 10^ 8 B^2$) and is thus used as standard for frequency
measurements, i.e. it is the line used in atomic clocks.
    
Unfortunately, only a (small) fraction of pairs will effectively be able to
create entanglement, because of large losses in transmission. However,
Alice and Bob can check which atoms have effectively received a photon
using a slightly more involved scheme. For example, after having generated
the superposition of the levels $6 ^2P_{3/2}$ $F=3$ $m_F= 1$ and $m_F=-1$,
they use a first Raman pulse for transferring the state $6 ^2P_{3/2}$ $F=3$
$m_F= 1$ into $ 6^2S_{1/2}$ $F=3$ $m_F= 1$ and the $6 ^2P_{3/2}$ $F=3$
$m_F= -1$ into $ 6^2S_{1/2}$ $F=3$ $m_F= -1$, which are stable. Then they
induce a transition from the level $6^2S_{1/2}$ $F=3$ $m_F=0$ and observe
if the atom fluoresces or not. They will then keep all the atom pairs,
where both the atoms did not fluoresce. 
Finally they use a second Raman pulse for transferring the level $
6^2S_{1/2}$ $F=3$ $m_F= 1$ into $ 6^2S_{1/2}$ $F=4$ $m_F= 0$ and  $
6^2S_{1/2}$ $F=3$ $m_F= -1$ into $ 6^2S_{1/2}$ $F=3$ $m_F= 0$, realising
the superposition of Eq. \ref{psi}.

It must also be emphasised that it is not important that the entangled
pairs are created exactly at the same time for one can stock them up to
when a sufficiently large ensemble is obtained and then start the clock by
Alice measurement.

The present status of the art of atom trapping is that atoms can be
confined for more than two minutes in a laser trap \cite{O'Hara}. This
lifetime can be increased to some hours housing the trap chamber in a
helium cryostat. Thus the trap time, and hence the decoherence time, is
already sufficient for implementing the entanglement procedure with
available technologies.
After the entanglement has been realised, the confining lasers must be
turned off during the clock measurement in order to eliminate light shift
effects. Anyway, with nowadays technologies, the atoms can be cooled enough
that, even after having turned off the confining lasers,  they remain
sufficiently confined for performing the clock measurement for some seconds
(however the effect of Raman transition on the cooling procedure should be
carefully investigated for realising a working apparatus.)

Of course, a similar scheme can be realised for any other clock based on
atomic transitions of trapped atoms and in particular for other alkaline
atoms as the rubidium (where one can use in the previous scheme the levels
$5 S_{1/2}, F=1,2, m_F=0$ and   $5 P_{1/2}, F=1,2, m_F=\pm 1$), which have
been proposed as frequency standard as well.
  
It must also be noticed that what is usually meant for synchronisation of
two clocks by standard frequency experts, it is not the starting of the
clocks at the same time, but the control of the frequency shift between them.
Let us now consider how a comparison between the frequencies of the two
clocks could be achieved using the previous scheme:

i) At a first instant $t_0=0$ Alice performs the Hadamard transform on a
sample of atoms correlated  with the Bob's ones. Using the method described
previously this procedure fixes a first common time $t_0=0$ shared between
Alice and Bob. 

ii) Alice and Bob measure their respective values of $\Omega$ ($\Omega_A$
and $\Omega_B$ respectively). 

iii) At a time $t_1=n/ \Omega_A$, where n is a sufficiently large known
number, Alice performs a Hadamard transform on a second sample. This allows
to synchronise this second pair of clocks at time $t_1$. $t_1=n/ \Omega_A$
is then a second time known both by Alice and Bob. Bob can then check how
much $t_1 \Omega_B$ differs by $n$, measuring the frequency shift of his
clock from Alice's one.  

In practice, Alice  can start the second clock when the first one is on a
maximum of $P_1(t)$, Bob will be able to deduce the frequency shift of his
clock respect to Alice's one just checking how much his second clock is
dephased respect to his first clock.

With this procedure the usual synchronisation between two atomic clocks is
thus performed.

Before concluding, let us also mention that the use of entanglement of more
photons of the form $\vert R R R ...\rangle + \vert L L L ...\rangle$ would
allow, using the scheme discussed in this paper, the creation of entangled
states of more atoms of the form $\vert 0 0 0 ... \rangle + \vert 1 1 1 ...
\rangle$, necessary for the set-up described in \cite{Chiara} for reducing
errors in frequency standard determination. The creation of states of three
polarisation entangled photons has already been obtained experimentally
\cite{Jian}, but with a very low efficiency. However, alternative schemes,
in principle suitable for higher efficiency and for entangling more
photons, have been proposed \cite{nos2}. As far as we know this represents
the only conceivable scheme, using available technologies, for realising an
entangled state of many cesium atoms. 

\vskip 0.3 cm
  
{\bf Acknowldgements}
\vskip 0.2cm
\noindent  Thanks are due to Aldo Godone for useful discussions.

We would like to acknowledge support of ASI under contract LONO 500172, of
Istituto Nazionale di Fisica Nucleare, of "Regione Piemonte" and of  MURST
via special programs "giovani ricercatori" Dip. Fisica Teorica Univ. Torino.


\begin{references}

\bibitem{Chiara} J.J. Bollinger et al, \PRA 54 (1996) R4649; S.F. Huelga et
al.,
Appl. Phys. B 67 (1998) 723.

\bibitem{Jozsa} R. Jozsa et al., \PRL 85 (2000) 2010.

\bibitem{lontano} W.T. Buttler et al., quant-ph 0001088; W. Tittel et al.,
quant-ph 9911109; H. Zbinden Appl. Phys. B 67 (1998) 743, W.T. Buttler et
al, Phys.Rev.Lett. 81 (1998) 3283 and ref.s therein.

\bibitem{BOTV} R. Bonifacio et al, Journ of Mod. Phys. 47 (2000) 2199,
Proc. of Quantum Mysteries, Garda 99, Ed. R. Bonifacio;  quant-ph 9906115.

\bibitem{paolo} P. Zanardi and M. Rasetti, \PRL 79 (1997) 3306.

\bibitem{viola} L. Viola, E. Knill and S. Lloyd, \PRL 82 (1999) 2417.

\bibitem{Ramsey1969}  N.F.~Ramsey, {\em Molecular Beams}, Oxford University
Press, Oxford, U.K., (1969), Sec. V.4. 

\bibitem{Audoin} C. Audoin, Proc. of "Enrico Fermi" school LXVIII, Varenna
(1980) 223.


\bibitem{seth} S. Lloyd et al., quant-ph 0003147.

\bibitem{nos} G. Brida, M.Genovese, C. Novero and E. Predazzi, Phys. Lett. A
268 (2000) 12; 
Proc. of QCM\&C 3, ed. O. Hirota and P. Tombesi (quant-ph/0009067), and
ref.s therin.

\bibitem{O'Hara} K. O'Hara et al., \PRL 82 (1999) 4204.

\bibitem{Jian} Jian-Wei Pan, D. Bouwmeester, M. Daniell, H. Weinfurter and
A. Zeilinger, Nature 403 (2000) 515.

\bibitem{nos2} see for example M. Genovese and C. Novero, \PLA 271 (2000) 48.

\end{references}
\end{document}